# The Role of Downward Momentum Transport in the Generation of Convective Downbursts

Ken Pryor

METO 658B

**Abstract**

A downburst index has been developed to assess the magnitude of convective downbursts associated with heavy precipitation-producing (HP) convective systems. The index, designed for use during the warm season over the central and eastern continental United States, is composed of relevant parameters that represent the simultaneous physical processes of convective updraft development and downburst generation, incorporating positive buoyant energy or convective available potential energy (CAPE) and the vertical equivalent potential temperature ($\theta_e$) gradient between the surface and the mid-troposphere. Since convective storm updrafts require buoyant energy, a very important parameter used in the analysis of convection is CAPE, computed from real-time Geostationary Operational Environmental Satellite (GOES) sounding data. Since updraft strength is proportional to CAPE, large CAPE would result in strong updrafts that could lift a precipitation core within a convective storm to a dry air layer in the mid-troposphere. CAPE also plays a major role in the formation of precipitation by resulting in strong updrafts that increase the size of precipitation particles, which, in turn, enhances the effect of precipitation loading. The amount of mid-level water vapor relative to the low-level water vapor, represented by the $\theta_e$ deficit ($\theta_{emax} - \theta_{emin}$) is important in the determination of downdraft strength due to evaporational cooling as dry air is entrained into the convective system. In addition to large CAPE and the presence of a vertical ($\theta_e$) gradient, previous research has identified other favorable conditions for downburst generation during cold-season convection events. Strong forcing by a vigorous upper-level troughs and rapidly moving cold-frontal systems as well as strong and diffluent winds aloft have been associated with severe wind-producing convective systems that develop during the cold season. A primary mechanism in downburst generation associated with synoptically forced convective systems is the downward transport of higher momentum possessed by winds in the mid-troposphere.

This paper will investigate the role of downward momentum transport in the production of convective winds observed on the Earth's surface. A literature review of the current understanding of momentum transport in convective systems as well as the types of convective systems that produce downbursts on a large scale will be presented. The paper will then feature case studies comparing a typical warm-season downburst-producing convection event to an early cold-season that occurred on 23 and 24 November 2004 in which it is believed that downward momentum transfer was a major factor in the magnitude of convective winds observed at the surface. Finally, based on a review of previous research and the case studies presented, a modification to the use of the downburst index will be explored to apply to cold-season forecasting situations.

# 1. Introduction

A multi-parameter index has been developed to assess the magnitude of convective downbursts associated with heavy precipitation-producing (HP) deep convective storm systems that occur over the central and eastern continental United States. A downburst is defined as a strong convective downdraft that results in an outburst of damaging winds on or near the earth's surface (Fujita and Wakimoto 1983). The Wet Microburst Severity Index (WMSI) is composed of relevant parameters that represent the simultaneous physical processes of convective updraft development and downburst generation, incorporating convective available potential energy (CAPE) and the vertical equivalent potential temperature ($\theta_e$) gradient between the surface and the mid-troposphere (Pryor and Ellrod 2004). Figure 1 illustrates the role of CAPE and $\Delta \theta_e$ in the process of updraft and downdraft formation, respectively. The WMSI algorithm is given as the following expression: WMSI = (CAPE)($\Delta \theta_e$)/1000, where $\Delta \theta_e = \theta_{emax} - \theta_{emin}$, where $\theta_{emax}$ refers to the maximum value of $\theta_e$ at the surface and $\theta_{emin}$ refers to the minimum value of theta-e in the midlevels of the troposphere (Atkins and Wakimoto 1991) . The Geostationary Operational Environmental Satellite (GOES) WMSI product ingests atmospheric sounding data (i.e., temperature and dew point) provided by satellite retrievals. Complete descriptions of the atmospheric sounding process are presented in Zehr et al. (1988) and Menzel et al. (1998).

Since deep convective storm updrafts require buoyant energy, a very important parameter used in the analysis of convection is CAPE, computed from real-time GOES sounding data. Since updraft strength is proportional to CAPE (Weisman and Klemp 1986), large CAPE would result in strong updrafts that could lift a precipitation core within a convective storm to a dry air layer in the mid-troposphere. CAPE also plays a major role in the formation of precipitation by resulting in strong updrafts that increase the size of precipitation particles, which, in turn, enhances the effect of precipitation loading in convective downdraft development (Doswell 2001; Wakimoto 2001). The amount of mid-level water vapor relative to the low-level water vapor, represented by the $\Delta \theta_e$ is important in the determination of downdraft strength due to evaporational cooling as dry air is entrained into the convective system (Atkins and Wakimoto 1991). The WMSI algorithm is designed for use during the warm season, from 1 May to 30 September. Pryor and Ellrod (2004) found that there exists a statistically significant correlation between GOES WMSI and the magnitude of observed surface wind gusts for both daytime ($r = 0.67$) and nighttime ($r = 0.65$) events during the warm season.

In addition to large CAPE and the presence of a vertical ($\theta_e$) gradient, previous research has identified other favorable conditions for downburst generation during cold-season deep convection events. Strong forcing by a vigorous upper-level troughs and rapidly moving cold-frontal systems as well as strong and diffluent winds aloft have been associated with downburst-producing convective systems that develop during the cold season. A primary mechanism in downburst generation associated with synoptically forced convective systems is the downward transport of higher momentum, possessed by winds in the mid-troposphere, into the planetary boundary layer.

This paper will investigate the role of downward momentum transport in the production of convective winds observed on the Earth's surface. A literature review of the current understanding of momentum transport in convective systems as well as the types of convective systems that produce downbursts on a large scale will be presented. The paper will then feature case studies comparing a typical warm-season downburst-producing convection event to an early cold-season that occurred on 23 and 24 November 2004 in which it is believed that downward momentum transfer was a major factor in the magnitude of convective winds observed at the surface. Finally, based on a review of previous research and the case studies presented, a modification to the use of the downburst index will be explored to apply to cold-season forecasting situations.

**2. Background**

Stull (1988) has noted that in a convective boundary layer momentum is typically transported upward, or up gradient, by updrafts associated with thermals. This results in a negative vertical momentum flux through the convective mixed layer. However, in a deep convective storm environment, this situation can be reversed with higher momentum air being transported to the surface as a result of intense convective mixing. Sasaki and Baxter (1986), in their analysis of convective storm morphology and dynamics, identified that downward transfer of entrained momentum from the strong environmental flow aloft was primarily responsible for the generation of strong surface winds. The authors note that descending evaporatively cooled air tends to carry the horizontal momentum that it had at its original level. Thus, strong winds aloft can be a significant predictive indicator of the potential for strong convective winds. Duke and Rogash (1992) continued this study by analyzing a severe, downburst-producing squall line that occurred in April 1991. This study also identified that downward transport of higher momentum, possessed by winds in the middle troposphere, by convective downdrafts was a major factor in the strength of surface outflow. Downward momentum transport is important when parcels, in an elevated dry (or low $\theta_e$) layer, conserve horizontal wind velocities as they become negatively buoyant and descend into the boundary layer. A significant additional finding was that wind directions associated with the surface convective wind gusts would suggest a contribution from downward momentum transfer. The authors contrasted this finding with the situation in which weak winds aloft were associated with considerable variability surface wind gust direction. Wakimoto and Bringi (1988) have noted that surface convective wind gusts, under those circumstances, radiate from the downburst impact area in a starburst pattern. Thus, the convective downburst process results in a temporary positive vertical momentum flux in the boundary (sub-cloud) layer, contrary to what is typically observed in the convective mixed layer.

This observational study will not quantify vertical momentum transport in the boundary layer in deep convective storm situations, but will serve to qualitatively explore the role of downward momentum transport in the magnitude of convective downbursts and express this finding in terms of the operational forecast process for predicting the magnitude of convective winds at the surface.

**3. Methodology**

Data from the GOES WMSI was collected during the period from 23 October to 24 November 2004 and validated against conventional surface data, as documented in Table 1. Measured wind gusts from surface weather observations, recorded during downburst events, were compared with adjacent WMSI values. In order to assess the predictive value of WMSI, GOES data used in validation were obtained for retrieval times one to three hours prior to the observed surface wind gust. WSR-88D base reflectivity imagery was utilized for each downburst event to verify that observed wind gusts were produced by convective systems. Particular radar reflectivity signatures, such as the bow echo and the weak echo channel (Fujita 1978; Przybylinski and Gery 1983), were effective indicators of the occurrence of downbursts. The type of convective system that produced each downburst was classified according to bow echo type as illustrated in Figure 2 and described in Przybylinski (1995). In addition, representative GOES sounding data (if available) or radiosonde observation (RAOB) data, nearest in time and space, was collected for each downburst event. GOES sounding data will typically not be available when a cloud deck is present in the field of view of the sounder. This results from the fact that cloud cover precludes the generation of a complete retrieval of temperature and moisture profiles. For each sounding, the height of the dry (low $\theta_e$) layer was documented as well as wind velocity and direction in the low $\theta_e$ layer. Correlation between observed surface and dry layer winds, and between GOES WMSI values and observed surface wind gust velocities was computed for the period 23 October to 24 November 2004.

**4. Case Studies**

*a. Warm Season Event: 7 June 2004 East Texas Downbursts*

A multi-cellular cluster of deep convection developed over east-central Texas, near College Station, during the afternoon of 7 June 2004. The air mass in which the convective activity was developing was statically unstable, due to intense solar heating of the surface, as displayed by Figure 3, the 1800 UTC GOES WMSI image. High WMSI values, as well as the presence of widespread towering cumulus convection, were an indicator of the strong instability in the region into which the convective cluster was propagating. Also apparent was the presence of a mid-tropospheric layer of dry (low $\theta_e$) air that could be entrained into the downdraft of a mature convective storm and result in subsequent downdraft acceleration and downburst development.

By 1913 UTC, NEXRAD (KSHV) reflectivity imagery animation (not shown) displayed the evolution of the convective cluster into a bow echo (Przybylinski 1995) in the region of strong instability that was characterized by towering cumulus and moderate WMSI values between 50 and 80. Downburst activity commenced upon development of the bow echo. The first observed downburst wind gust of 40 knots occurred at Palestine at 1925 UTC, where a well-defined bow echo was indicated in radar reflectivity imagery. The bow echo continued to track to the northeast during the next hour into a progressively more unstable air mass with increasing WMSI values. At 2011, a stronger downburst of 55 knots was observed at Tyler, where a considerably higher WMSI value of 172 was indicated. Accordingly, as displayed in Figure 4, a distinctive bow echo was indicated by radar imagery in the vicinity of Tyler at the time of downburst occurrence.

In this case, large positive buoyant energy resulted in strong updrafts that lifted the precipitation core within the convective storms to the mid-level dry air layer. Large WMSI values implied the presence of large convective available potential energy (CAPE) as well as relatively dry air at mid levels that would result in evaporative cooling and the generation of large negative buoyancy as dry air was entrained into the convection cells (Pryor and Ellrod 2004). Figure 5, the 1200 UTC radiosonde observation (RAOB) from nearby Shreveport, Louisiana displayed conditions typically expected over the southeastern United States during the summer season. Especially noteworthy is the presence of only weak unidirectional wind shear in the low and middle levels of the troposphere, up to about 500 mb, with very weak winds above the 500-mb level. In fact, wind velocities in the low $\theta_e$ layer, located between the 600 and 700-mb levels, were only around 20 knots. Compared to the magnitude of the downburst wind gusts observed at the surface, it is apparent that the role of downward momentum transport in this case was minimal. Therefore, in this typical warm season event, buoyancy and instability effects primarily drove downburst strength with negligible contribution from downward momentum transport.

*b. Cold Season Event: 23-24 November 2004 Severe Squall Line*

During the evening of 23 November 2004, a squall line (convective line) developed along a strong cold front over central Texas ahead of negatively tilted upper-level short wave trough. The squall line then tracked eastward into a highly unstable air mass during the evening and overnight hours, producing several severe downbursts and wind damage over eastern Texas. A peak convective wind gust of 52 knots was observed at Galveston, where WMSI values in excess of 100 had been previously indicated. Pryor and Ellrod (2004) found that WMSI values between 80 and 200 are correlated with surface wind gusts of 50 to 64 knots. In addition, the pre-convective environment was characterized by strong wind shear in the low and middle levels of the troposphere. The downward transport, by the strong convective downdrafts, of higher momentum from the mid-troposphere to the surface most likely resulted in the strong winds produced by the squall line. Observed surface wind gusts associated with this convective system and corresponding GOES WMSI values are noted in Table 1.

In Figure 6a, 0000 UTC 24 November 2004 GOES WMSI product, enhanced infrared (IR) satellite imagery displays a convective line developing along a cold front, ahead of a negatively-tilted short wave trough (indicated by red dashed line). The cold front was apparent in the IR imagery as a low-level temperature gradient. The air mass ahead of the developing squall line was convectively unstable and favorable for downbursts as indicated by the large region of WMSI values in excess of 80 over southeastern Texas. Strong surface convergence along the cold front and positive vorticity advection (PVA) ahead of the short wave trough enhanced lifting and served as an initiating mechanism for deep convection. In addition, the negatively tilted trough, sloping in the direction opposite to the upper-level wind flow with latitude, provided favorable conditions for deep convective-storm activity by enhancing vertical circulation and static instability. In this image, a well-defined comma cloud signified the presence of the short wave trough. Deep convection was developing in the comma cloud "tail" where the environment was most potentially unstable.

Over the next three hours, the convective line intensified as it moved eastward. By 0200 UTC (Figure 6b), overshooting tops were apparent, indicated by the cold temperatures in the enhanced IR imagery, where convective updrafts were penetrating the cloud tops. Accordingly, radar reflectivity imagery animation (not shown) from the Houston, Texas WSR-88D, displayed that the squall line had evolved into a type 1 bow echo pattern (Przybylinski 1995) with several bowing line segments signifying the location of strong downbursts. Significant downburst winds were observed at College

Station and Victoria between 0200 and 0300 UTC, corresponding to the location of bow echoes embedded along the squall line (Figure 7a). Between 0300 and 0500 UTC, the squall line continued to propagate eastward, producing several strong downbursts in the Houston metropolitan area that were effectively portrayed by satellite and radar reflectivity imagery (Figures 6 and 7). Shortly before the squall line moved over the Gulf of Mexico, the strongest observed downburst wind gust of 52 knots was recorded in Galveston at 0500 UTC, adjacent to the highest WMSI values in the region. In a similar manner to the previously observed downbursts, the severe convective wind gust was associated with a well-defined bow echo as identified in radar reflectivity imagery.

      In contrast to the warm season case, this downburst event was the result of strong synoptic-scale forcing in the presence of extreme convective instability and a positively buoyant air mass. Also, as expected with this cold season case, strong vertical wind shear was in place with high winds and large momentum present in the mid-tropospheric dry air layer, as was most effectively portrayed in Figure 8, the GOES soundings from Victoria at 0100 UTC and South Brazos (the closest representative sounding retrieval location to Galveston) at 0400 UTC, respectively. The channeling of low $\theta_e$ air into the rear of each individual bowing line segment was indicated by the presence of rear inflow notches (RINs), as displayed in Figure 7. The entrainment of low $\theta_e$ air into the convective system resulted in evaporative cooling and the subsequent generation of negative buoyancy and downdraft acceleration. The intense downdrafts transported the higher momentum air from the midlevels of the troposphere to the surface, as was evidenced by the close correspondence between the velocity and direction of the mid-tropospheric (low $\theta_e$ layer) winds and wind gusts observed at the surface. This event

demonstrated the importance of downward momentum transfer in the magnitude of downburst wind gusts associated with cold-season convective systems.

## 5. Discussion

Table 1 documents the location, time, magnitude and direction of 20 downburst events recorded during the period 23 October to 24 November 2004. Also listed in the table are the corresponding WMSI values, dry layer winds, and the type of convective system that generated each downburst. As expected during the cold season, downbursts were produced by highly organized, linear convective systems. Computed correlation between observed surface and dry layer wind velocity and direction was 0.71 and 0.85, respectively, indicating a strong relationship between surface and mid-tropospheric dry layer winds associated with these downburst events. However, the correlation between WMSI values and surface wind gust magnitude was much weaker, only computed to be 0.32. Based on previous research and the case studies presented in this paper, the strong correlation between mid-tropospheric and surface winds suggested that downward transport of momentum from the midlevels of a convective storm environment to the surface played a major role in the downburst magnitude during this observation period.

Also notable is the variability in the altitude of the dry air layer for these downburst events. The dry air layer was typically located at a higher level in the troposphere for the southern events (TX, OK, MS) that what was observed for the northern events (WI, MN, IA). This raises an important forecasting issue in the utility of the GOES WMSI during cold season events. Efficient downward momentum transfer occurred when the convective updrafts were strong enough to lift the precipitation core to the level of the dry air layer. As apparent for the southern events, where the dry air layer was typically located between the 500 and 700-mb levels, larger WMSI values were required to result in momentum transfer between the mid-troposphere and surface. However, for the northern events, where the dry air layer was typically located between the 750 and 800-mb levels, only modest WMSI values, between 28 and 66, were associated with downburst wind gusts. This underscores the importance of considering the level of the mid-tropospheric dry layer when using the GOES WMSI product to forecast, in the short term, or "nowcast" the magnitude of convective downbursts. Unlike during warm season events that are typically characterized by strong static instability and weak vertical wind shear, the absolute value of the WMSI is arbitrary during synoptically forced, cold season events. The significance of the WMSI is in its relation to the level of the dry air layer and the likelihood that instability is sufficient to result in updrafts that will lift the precipitation core to the mid-levels of the convective storm, whereby entrainment will occur and result in downdraft acceleration.

**6. Summary and Conclusions**

This study has identified an important relationship between GOES WMSI and downward momentum transport in cold-season deep convective storms. Based on a review of previous literature, and an analysis of real-time surface observations, GOES WMSI product imagery and GOES sounding data for 20 cold-season downburst events, it has been found that downward transport of higher momentum, possessed by winds in the mid-troposphere, into the boundary layer, is a major factor in downburst magnitude. This finding was exemplified by two case studies that contrasted warm and cold-season downburst events and thus, highlighted the role of downward momentum transport in the strength of convective wind gusts observed at the surface. The bow echo and associated rear-inflow notch signature, as identified in radar reflectivity imagery, served to illustrate the physical process that resulted in downward momentum transfer. A strong correlation between surface and mid-tropospheric dry layer winds emphasized further the interconnection between downward momentum transfer and downburst magnitude. Also important is the relationship found between WMSI values and the level of the low $\theta_e$ layer, where higher momentum is entrained into a convective system and is eventually transferred to the surface. This study underscores the relevance of considering the level of the mid-tropospheric dry layer when using the GOES WMSI product to "nowcast" the magnitude of convective downbursts.

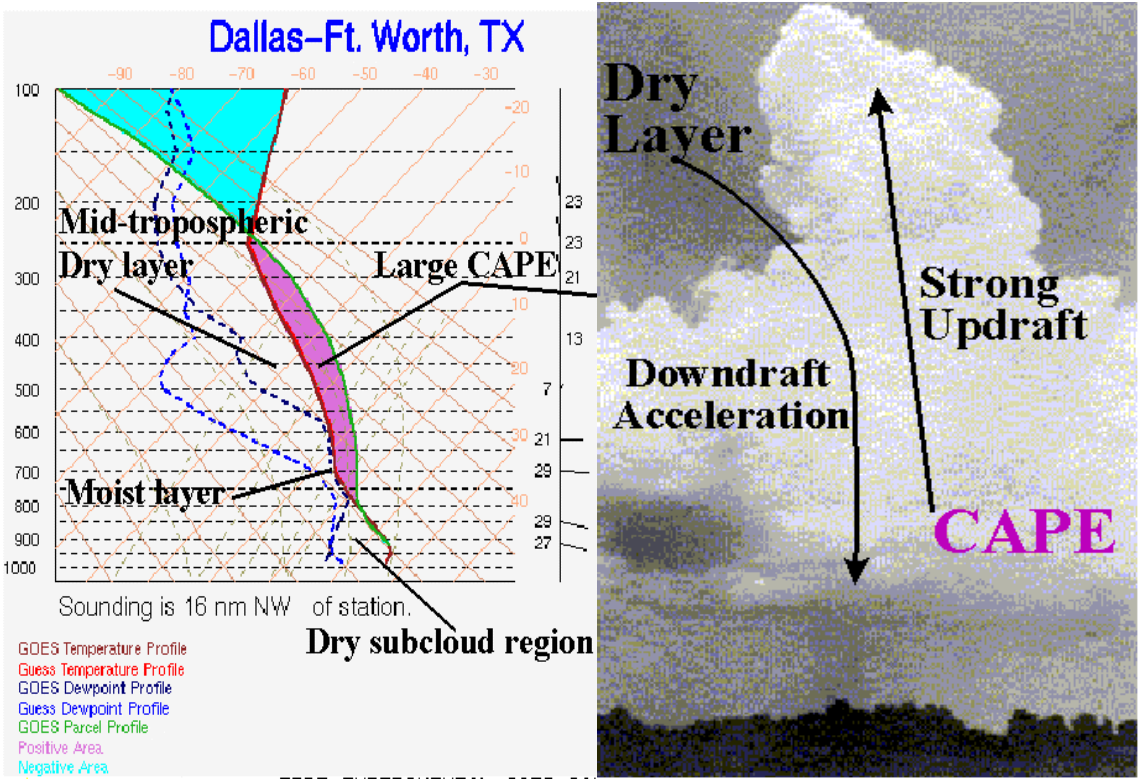

Figure 1. Comparison of a GOES derived sounding to a photograph of a single cell convective storm that produced a wet microburst (Atkins and Wakimoto 1991).

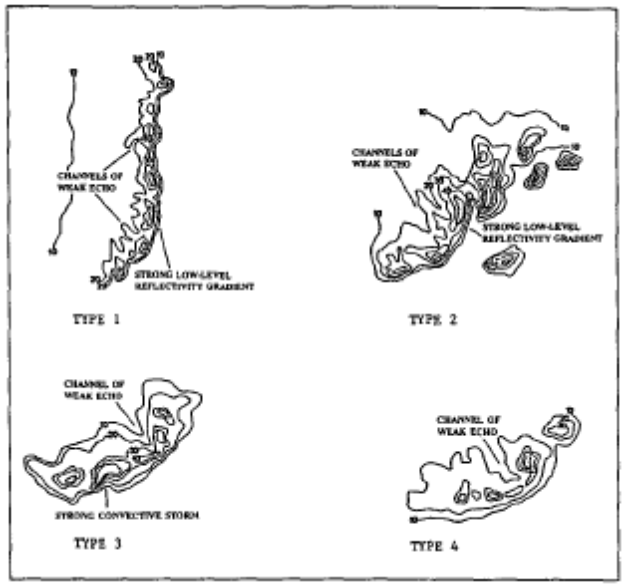

Figure 2. Schematic diagrams of bow echo radar signatures (Przybylinski 1995).

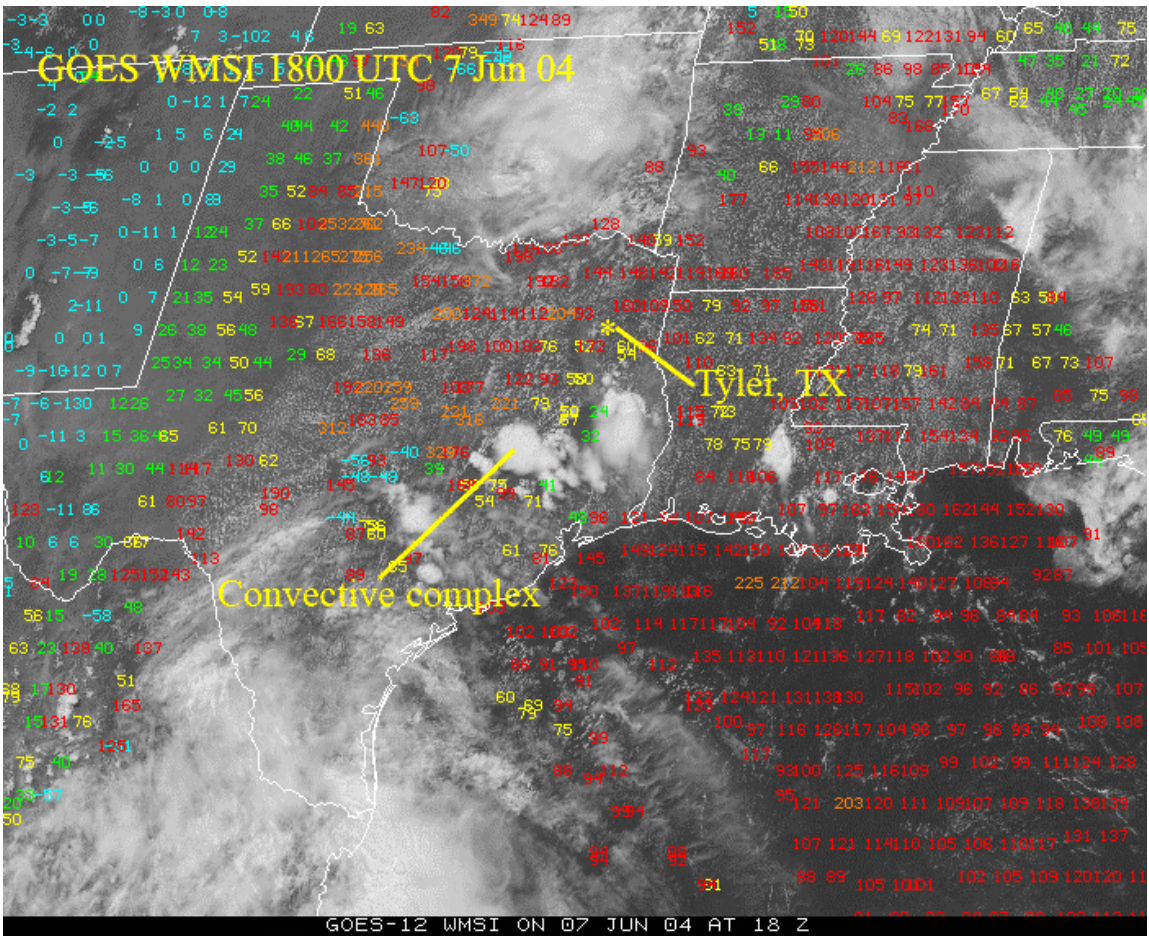

Figure 3. 1800 UTC 7 June 2004 GOES WMSI image.

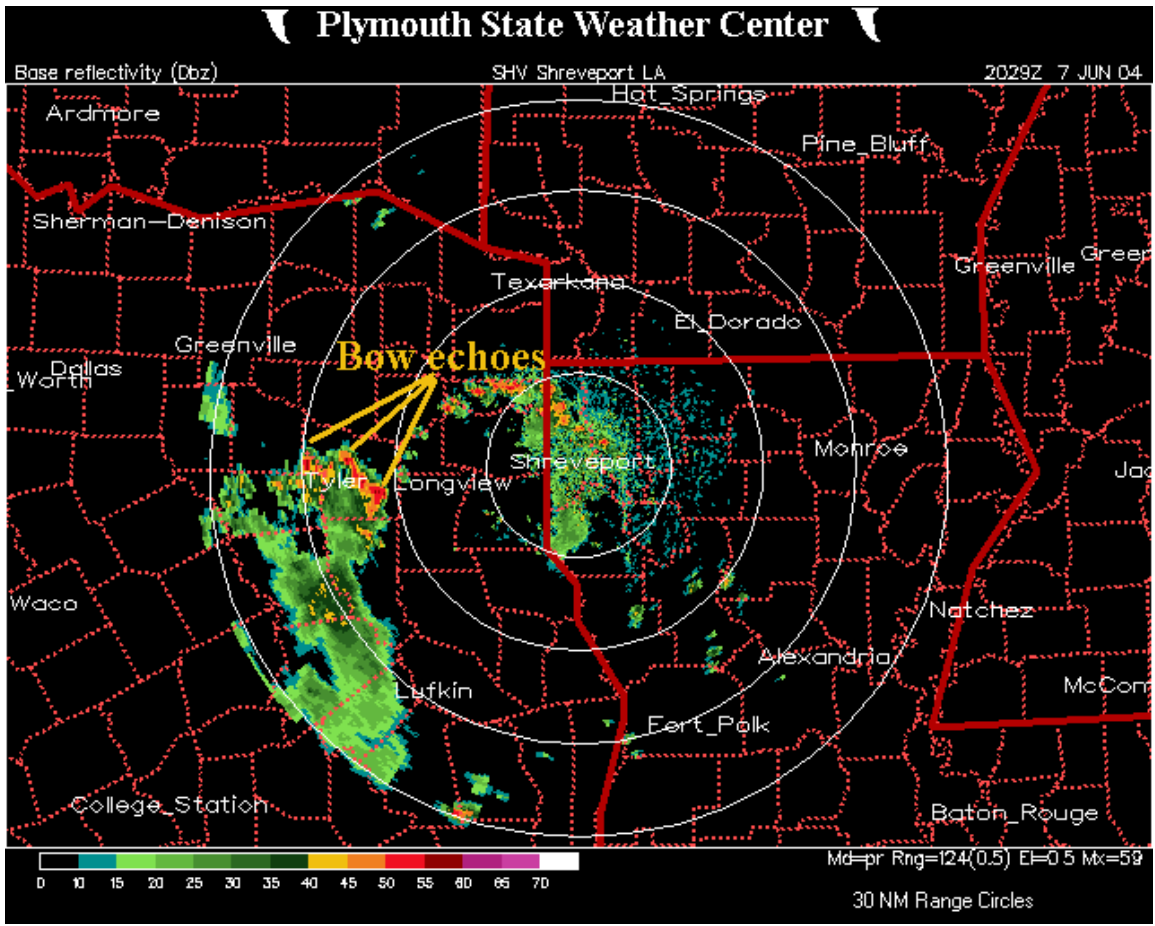

Figure 4.  2029 UTC 7 June 2004 NEXRAD (KSHV) reflectivity image.

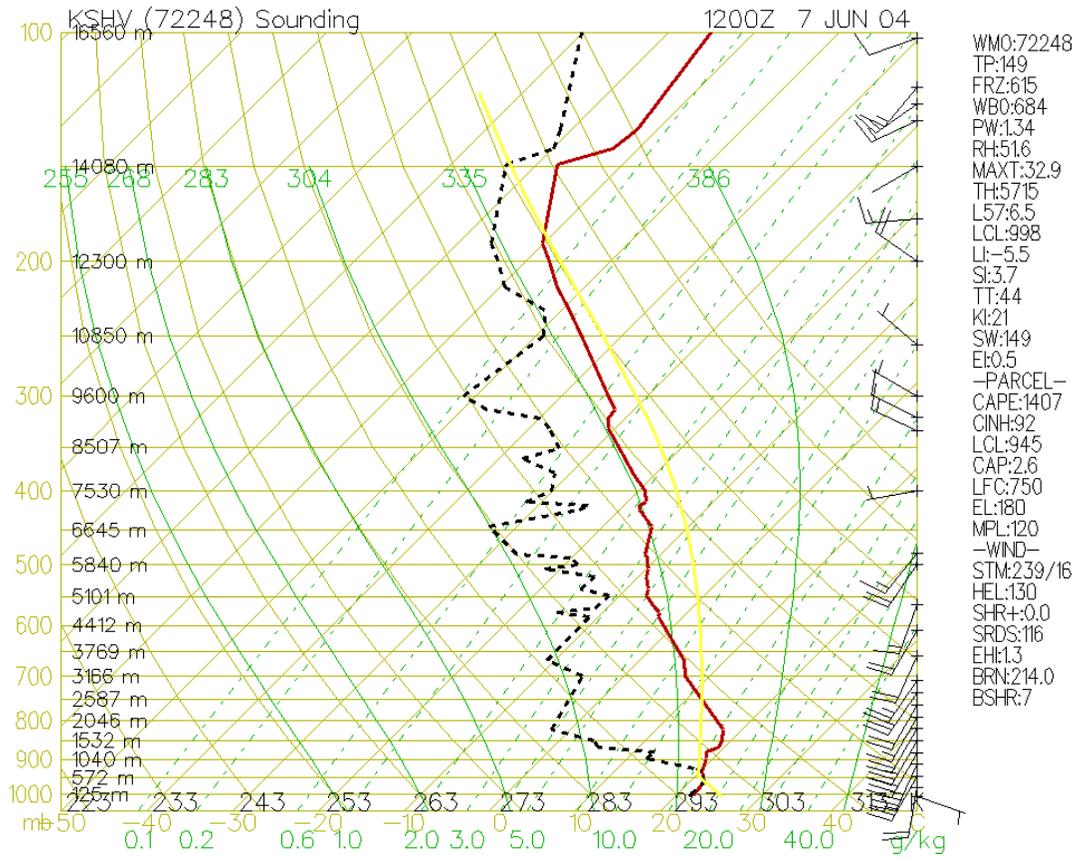

Figure 5. 1200 UTC 7 June 2004 radiosonde observation (RAOB) from Shreveport, Louisiana.

(a)

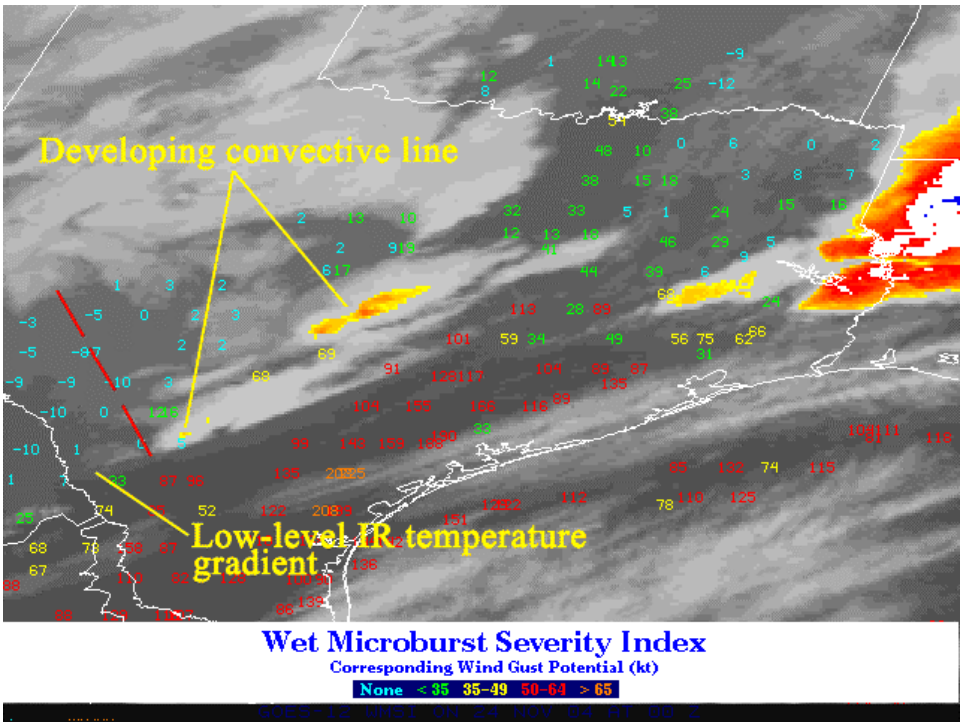

(b)

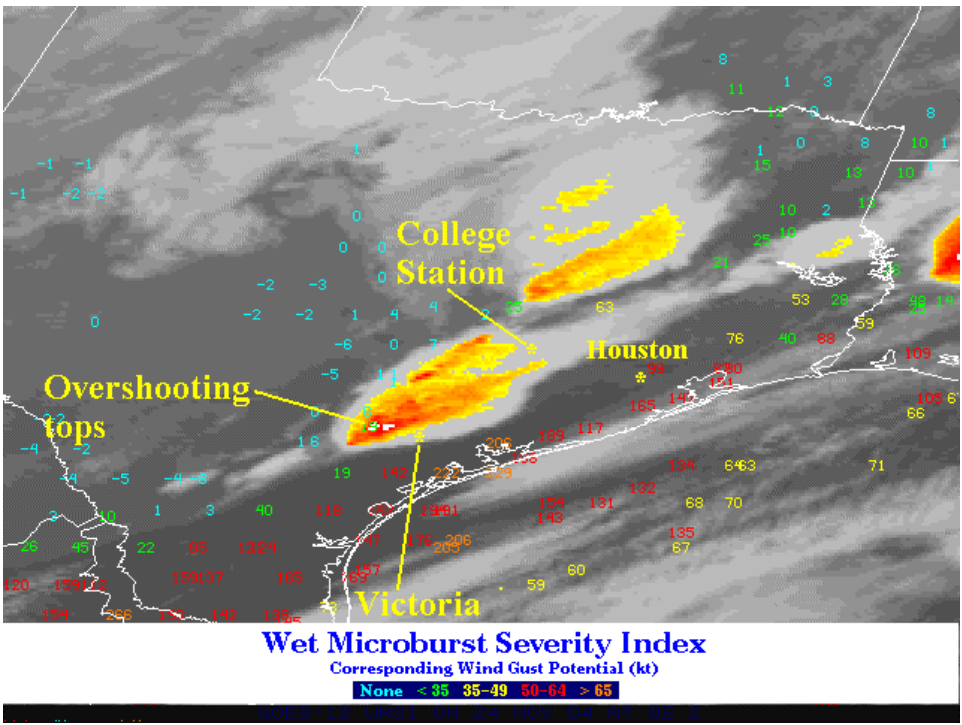

(c)

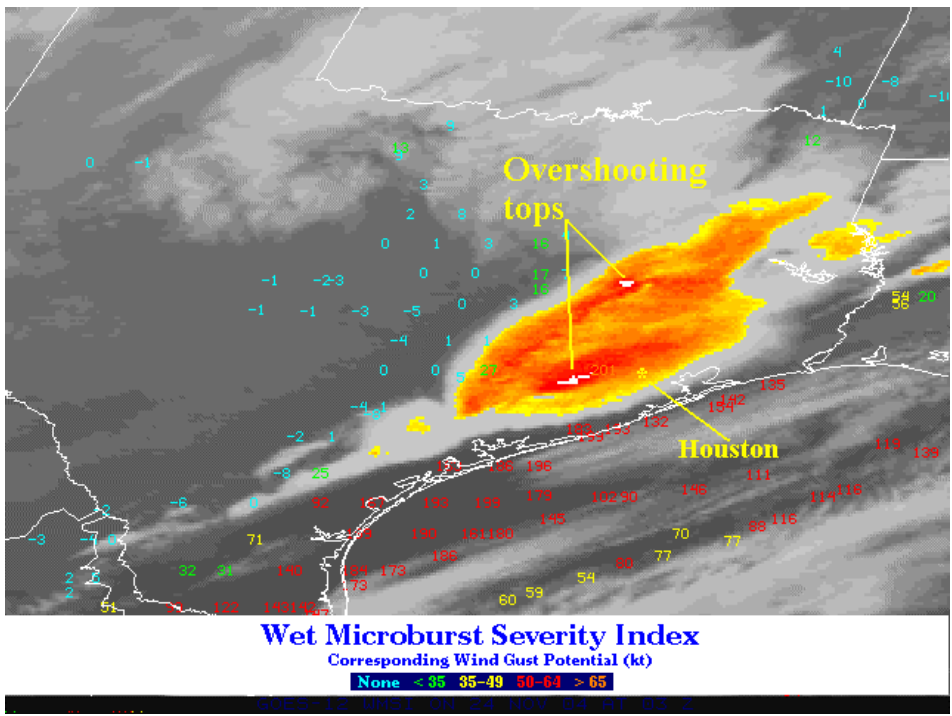

(d)

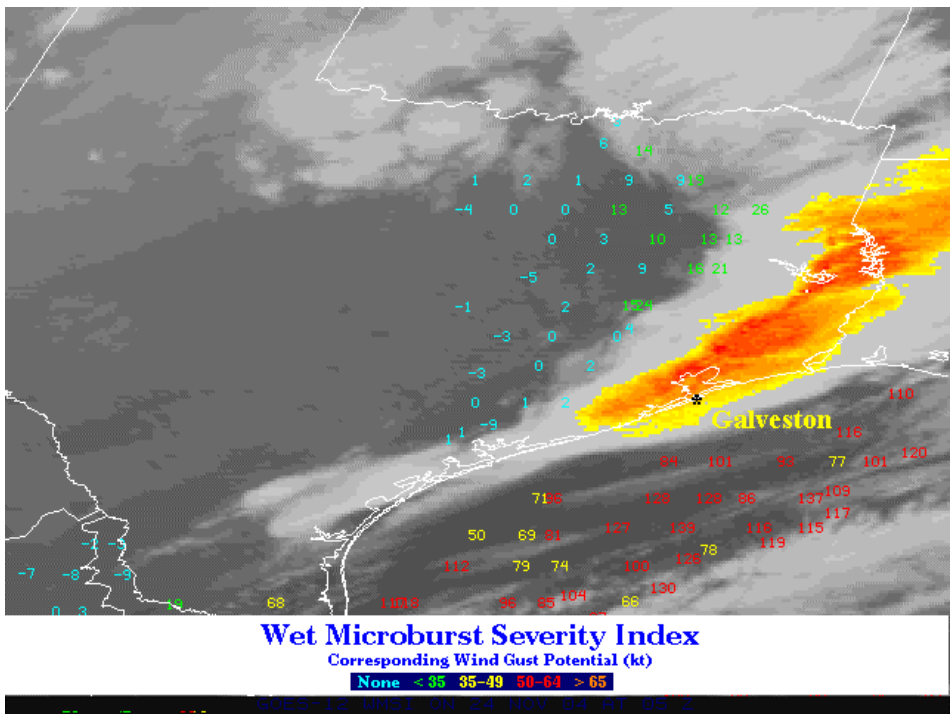

Figure 6. GOES WMSI imagery from 24 November 2004 at a) 0000 UTC; b) 0200 UTC; c) 0300 UTC; d) 0500 UTC.

(a)

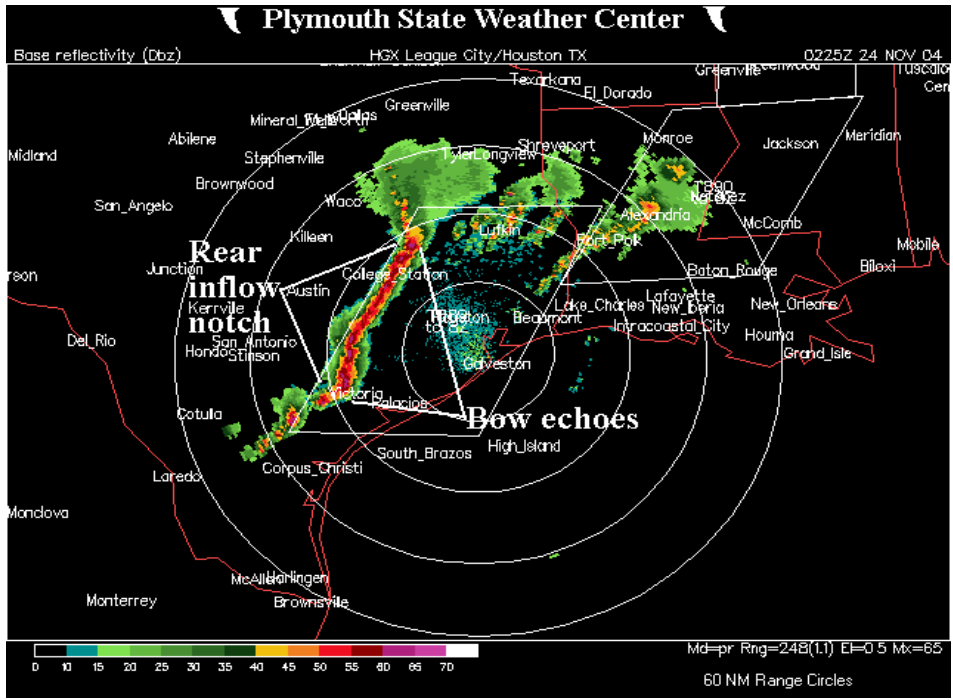

(b)

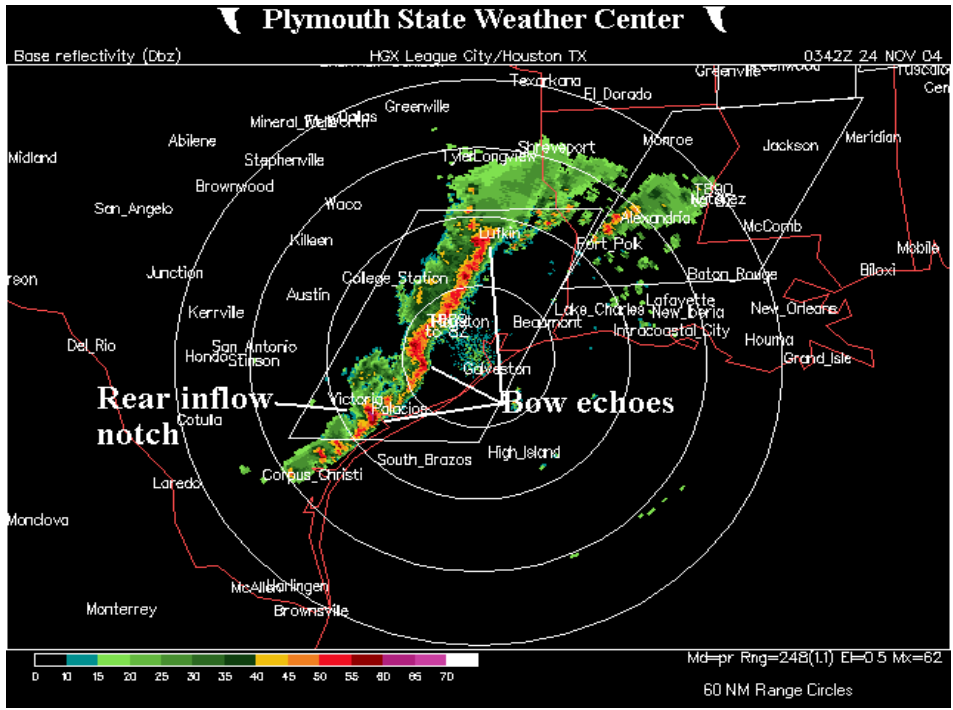

(c)

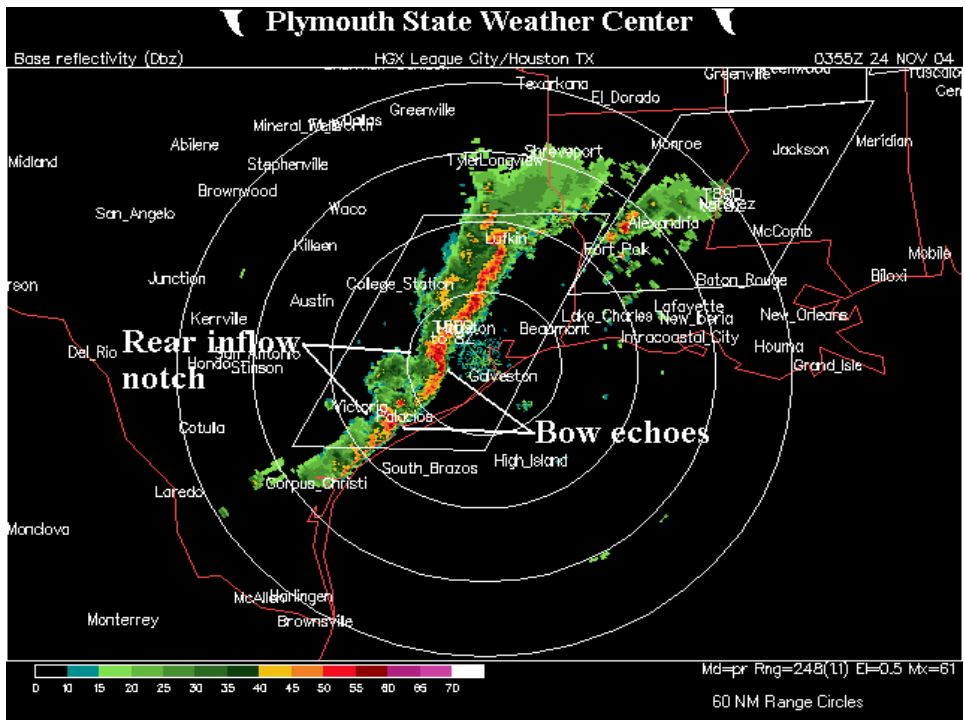

(d)

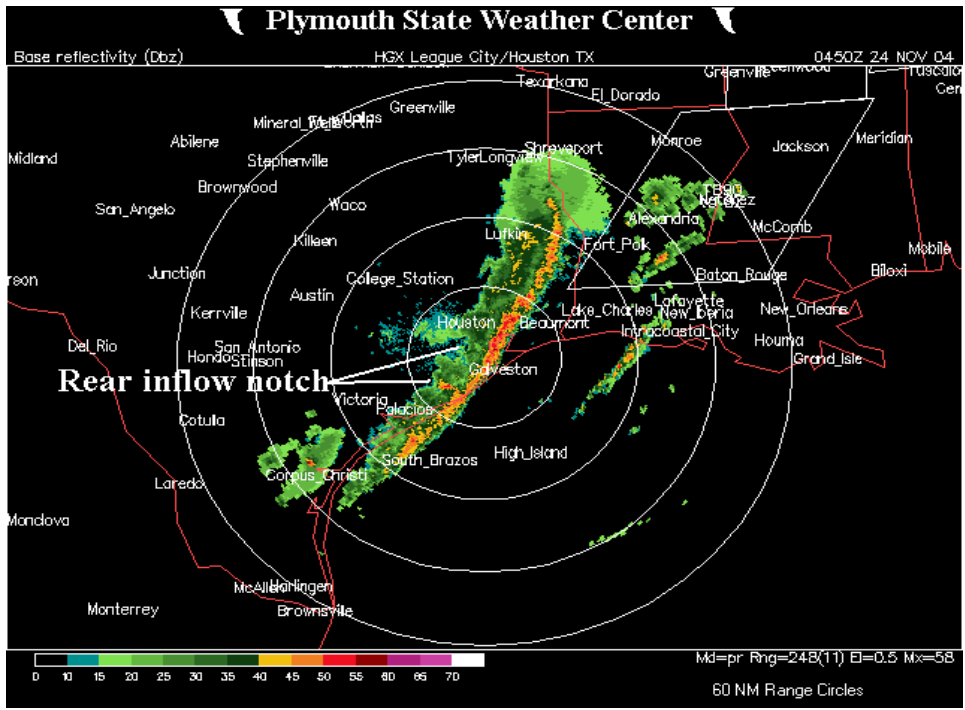

Figure 7. NEXRAD (KSHV) reflectivity imagery from 24 November 2004 at a) 0225 UTC; b) 0342 UTC; c) 0355 UTC; d) 0450 UTC.

(a)

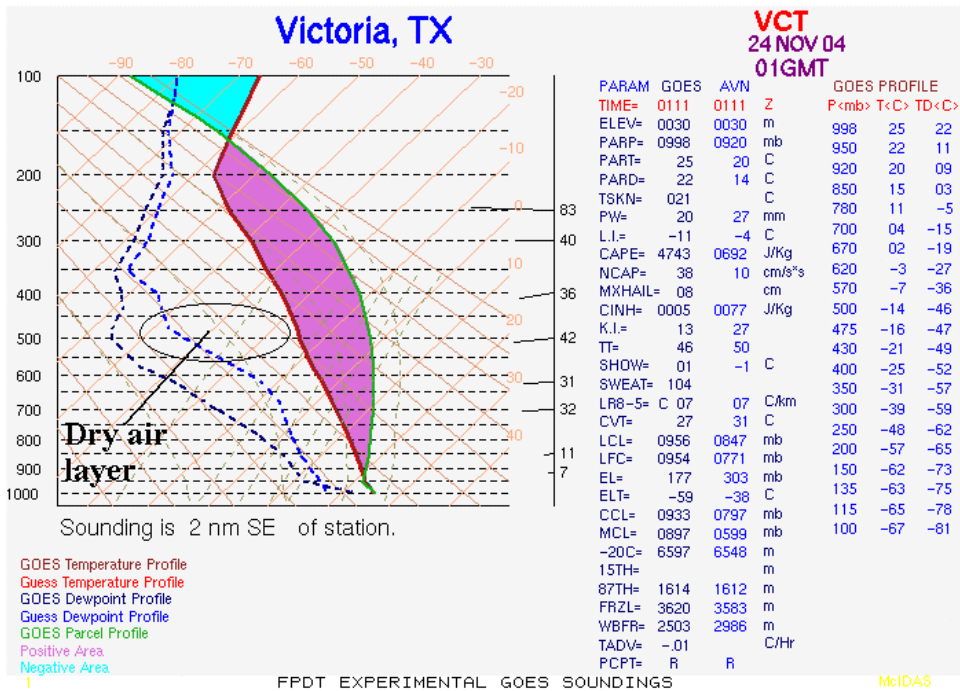

(b)

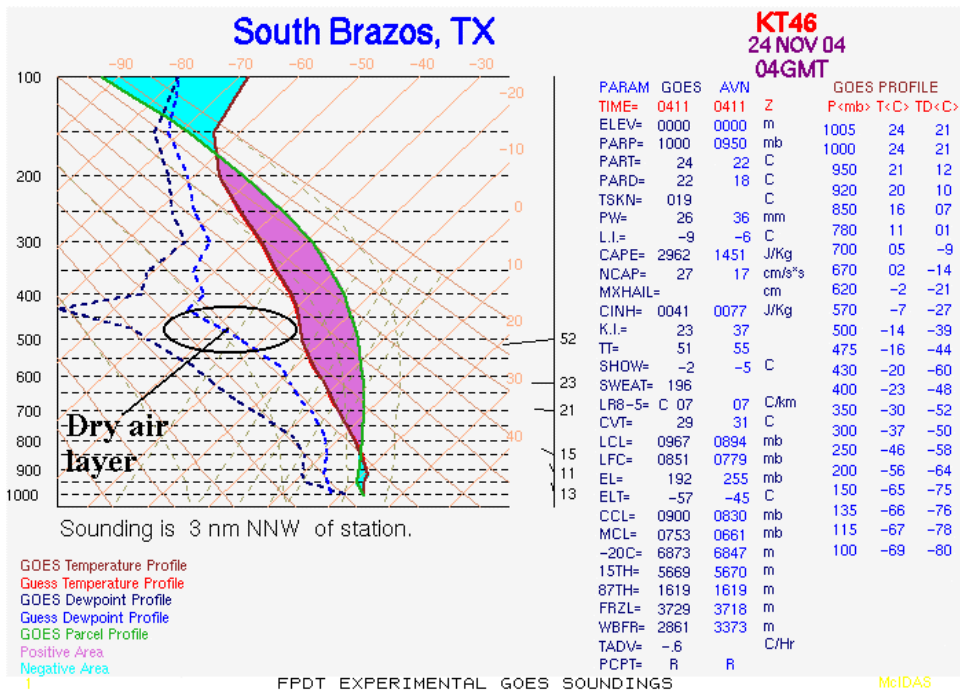

Figure 8. GOES soundings for 24 November 2004 at a) Victoria, Texas, 0100 UTC; b) South Brazos, Texas, 0400 UTC.